# Higher order Conjugate Exceptional Points in an 1D Photonic Bandgap Waveguide

**Sibnath Dey[1*], Harish N S Krishnamoorthy[1] and Somnath Ghosh[2]**
[1]*Tata institute of Fundamental Research, Hyderabad, pin-500046, India and Mahindra University Hyderabad, Pin-500043, India*
*Author e-mail address: **sibnath.dey1991@gmail.com**[*]*

**Abstract:** We demonstrate third-order conjugate exceptional points ($EPs$) in a gain-loss assisted multi-mode $1D$ complementary photonic bandgap waveguide. Our study reveals the higher-order mode conversion phenomenon facilitated by parametrically encircled third-order conjugate $EPs$, showcasing the potential for on-chip mode conversion.

## 1. Introduction:

One of the distinct non-Hermitian characteristics in various quantum-inspired and photonic systems is the occurrence of exceptional points ($EPs$) of different orders [1, 2]. At an $EP$, the coupled eigenvalues and corresponding eigenvectors merge, rendering the system Hamiltonian defective [1-4]. Encircling the system parameters quasi-statically around an $EP$ facilitates adiabatic state flipping among the system's eigenvalues, leading to non-adiabatic behavior in the time-asymmetric dynamics of eigenstates during the dynamical parametric encirclement process. Recently, numerous open photonic systems featuring different order $EPs$ have gained significant attention, particularly in photonic structures such as waveguides, microcavities, lasers, and photonic crystals. These systems have demonstrated a variety of intriguing applications, including asymmetric mode conversion, topological state flipping, anti-lasing, mode-selective optical isolation, and highly sensitive detection [1, 2]. The theoretical concept of different-order conjugate $EPs$ can be understood through the analysis of planar waveguide structures [3, 4]. In this study, we use a multi-mode supported **$1D$** photonic bandgap waveguide (**$PBG$**) framework to implement two complementary active perturbations. The chosen geometry is in the form of an unbalanced multi-core gain-loss profile with different widths, leading to the hosting and identification of third-order conjugate **$EPs$** (**$EP3$ and $EP3^*$**). We identified these third-order conjugate **$EPs$** through the simultaneous presence of two pairs of second-order conjugate **$EPs$** among three coupled **$TE$** modes in two complementary variants of **$1D$** waveguides.

## 2. Results and discussions:

We propose the design of a $1D$ $PBG$ waveguide featuring three bi-layers claddings in each arm, as illustrated schematically in Fig. 1(a). The periodic layers have refractive indices of $n_1 = 2.15$ and $n_2 = 3.927$, respectively. The widths of the layers with higher and lower refractive indices are 0.151 $\mu m$ ($d_1$) and 0.15 $\mu m$ ($d_2$), respectively. The transverse width ($W$) of the waveguide is 7.806 $\mu m$, with an optimized propagation length ($L$) of 5 mm. The core has a refractive index $n_{co}$ of 1.450 and a width ($d_{co}$) of 6 $\mu m$. The chosen operating wavelength ($\lambda$) is 1.625 $\mu m$. Non-Hermiticity is introduced through an unbalanced multilayer gain-loss profile (i.e. six alternating different width layers of gain and loss) in the core region, characterized by the gain coefficient ($\gamma$) and the loss-to-gain ratio ($\tau$). We consider two complementary types of active pumping, $WG_a$ and $WG_c$, within the same passive waveguide, ensuring a $T$-symmetric complex potential. This complex transverse refractive index profile is depicted in Fig. 1(b). The complex refractive indices within the gain and high-loss regions of the proposed waveguide core ($WG_a$) can be expressed as ($n_{co} - i\gamma$) and ($n_{co} + i\tau\gamma$) respectively. Similarly, the complex refractive indices within the loss and high-gain regions of the complementary waveguide core ($WG_c$) can be expressed as ($n_{co} + i\gamma$) and ($n_{co} - i\tau\gamma$). The waveguides exhibit two distinct complex refractive index profiles, $n_a$ and $n_c$ which are complementary due to their opposite gain-loss distributions along the transverse ($x$) axis. The waveguide supports five $TE$ modes, which are coupled with introducing non-Hermiticity through the gain-loss parameter. In Fig. 1(c), we present the three quasi-guided $TE$ ($TE_0, TE_1$ and $TE_3$) modes. By tuning the gain-loss profile, we identified two connected pairs of conjugate second-order exceptional points ($EP2s$) ($EP2^1, EP2^{1*}$ and $EP2^2, EP2^{2*}$) in the two complementary waveguides among these three interacting modes, as shown in Fig. 1(d). Accordingly, for both systems, we choose the specific range of $\gamma$ within [0, 0.2], and also choose different $\tau$-values and observed that $n_{eff0}$ and $n_{eff1}$ coalesce and exhibit an $EP2$ ($EP2^1$ & $EP2^{1*}$) at (0.112, 1.5) in the ($\gamma, \tau$)-plane, as shown in red and black curve in Fig.1 (d). Similarly, the $n_{eff1}$ and $n_{eff3}$ exhibit other pairs of conjugate EP2 (say: $EP2^2$ & $EP2^{2*}$) at (0.05, 3.935) ($\gamma, \tau$)-plane, as shown in same plot. As can be observed in Fig.1 (d), for $EP2^1$ the coalescence is occurring along the positive $Im(n_{eff})$ axis for $WG_a$ as it is loss dominated; however, for $EP2^{1*}$ it is observed along the negative $Im(n_{eff})$ axis for $WG_c$ as it is gain dominated. However, at $EP2^2$, we observed that coalescence occurs between $n_{eff1}$ and $n_{eff3}$ in positive

$Im(n_{eff})$ for $WG_a$ and we also observed that for $EP2^{2*}$ coalescence occurs in negative $Im(n_{eff})$ (for $WG_c$), as shown in black and green curve in the same Figure. Thus, employing two $T$-symmetric gain-loss profiles in a single passive waveguide, we have identified two pairs of conjugate $EP2s$ within the parameter space. We investigate the behavior of third-order branch point behavior by simultaneously encircling two pairs of conjugate $EP2s$ in two complementary active geometries. To investigate this behavior, we have chosen the parametric equations $\gamma(\phi) = \gamma_0 \sin(\phi/2)$ and $\tau(\phi) = \tau_0 + r \sin(\phi)$, where $r$ is a characteristic parameter, and $\phi$ is a tunable angle within $[0, 2\pi]$ that governs the closed variation of $\gamma$ and $\tau$. Here, we have chosen $\gamma_0 = 0.14$, $\tau_0 = 3.1$, and $r = 2$ to properly encircle both $EP2s$ simultaneously. This encirclement process is illustrated in Fig. 1(e), where the shape of the parameter space ensures that $\gamma = 0$ at both the beginning and the end of the encirclement. Now, we track the trajectories of the selected three $n_{eff}$-values in the complex $n_{eff}$-plane, as depicted in Fig. 1(f), and by following the stroboscopic variation of $\gamma$ and $\tau$ along the specified loops shown in Fig. 1(e). The trajectories of $n_{effj}$ ($j = 0,1,3$) in the complex $n_{eff}$- plane for two complementary variants are presented in the clockwise direction. We observed that, over one complete loop in the parameter space, all three coupled modes exchange their identities ($n_{eff0} \rightarrow n_{eff1} \rightarrow n_{eff3} \rightarrow n_{eff0}$). This unconventional dynamics of the three coupled modes confirms the presence of a pair of conjugate $EP3s$ in the system parameter space. The discovery of these mode dynamics, enriched with the physics of higher-order conjugate $EPs$, presents a promising platform for exploring the optical responses of two $T$-symmetric systems. This could significantly advance light manipulation techniques in integrated device applications.

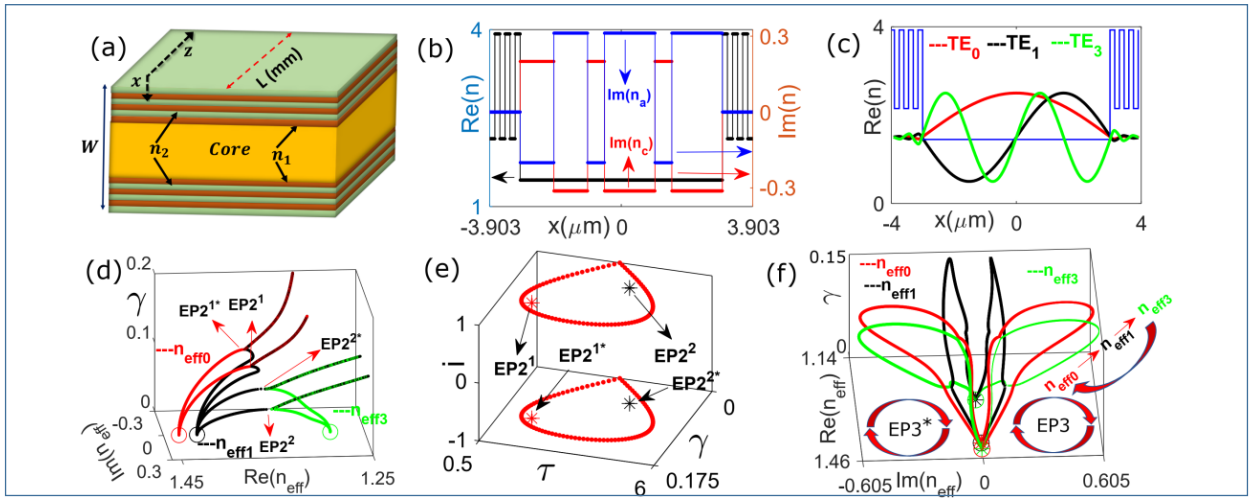

**Fig. 1:** *(a)* Diagram of a photonic bandgap waveguide (x as the transverse axis and z as the propagation axis) illustrating two $T$-symmetric gain-loss assisted complementary systems ($WG_a$ & $WG_c$). *(b)* Transverse complex refractive index profiles of two complementary systems. The real part of the refractive index [$R(n)$] is shown by the black line, while the imaginary part [$Im(n)$] is depicted by the red and blue dotted lines for $\gamma = 0.114$ and $\tau = 1.5$, respectively, corresponding to $WG_a$ and $WG_c$ on the right vertical axis. *(c)* Real part of the refractive index and field amplitude profiles of three quasi-guided TE modes $TE_j$ ($j = 0, 1, 3$). *(d)* Coalescence of three complex $n'_{eff}s$ values ($n_{eff0}, n_{eff1}, n_{eff3}$) at two different locations in two complementary systems: at $\gamma = 0.114$ for a chosen $\tau = 1.5$, for $EP2^1$, and at $\gamma = 0.0506$ for a chosen $\tau = 3.9375$, for $EP2^2$. *(e)* Two different selected parameter spaces in the $(\gamma, \tau)$−plane to encircle conjugate $EP3$, associated with $T$-symmetric complementary variants, as shown with the additional $i$-axis. *(f)* Third-order successive mode switching in the complex $n_{eff}$-plane for two conjugate variants of the waveguide, following clockwise parametric encirclements by the loops shown in *(e)*.

### 3. Summary:

In summary, we have hosted third-order conjugate $EPs$ in two complementary active $1D$ photonic waveguide systems, utilizing the framework of a common passive waveguide. These two waveguide variants exhibit unbalanced gain-loss distributions, resulting in complex conjugate refractive index profiles that are correlated by $T$-symmetry. We have demonstrated the robust branch point characteristics of the embedded conjugate higher order $EP$ under parametric encirclement, based on enclosing $EP2s$. This work, enriched with the physics of higher-order conjugate $EPs$, will open up a new paradigm for unconventional on chip light manipulation.

### 4. References:

1. Miri M-A and Alú A, "Exceptional points in optics and photonics" Science **363**, 6422 (2019).
2. E. J. Bergholtz, J. C. Budich, and F. K. Kunst, "Exceptional topology of non-Hermitian systems," Rev. Mod. Phys. **93**, 015005 (2021).
3. A. Laha, S. Dey, and S. Ghosh, "Reverse-chiral response of two T −symmetric optical systems hosting conjugate exceptional points" Phys. Rev. A **105**, 022203 (2022).
4. S. Dey, and S. Ghosh "Anomalous nonchiral light transport in the presence of local nonlinearity around higher-order conjugate exceptional points" Phy. Rev A. **108**, 023508 (2023).